\documentclass[aps,prl,reprint,superscriptaddress,preprintnumber,showpacs,twocolumn]{revtex4}

\usepackage{graphicx}
\usepackage{hyperref}
\usepackage{color}
\usepackage{dcolumn}
\begin{document}

\title{Symmetry-enforced Fermi degeneracy in  topological semimetal RhSb$_3$}

\author{Kefeng Wang}
\author{Limin Wang}
\author{I-Lin Liu}
\affiliation{Maryland Quantum Materials Center, Department of Physics, University of Maryland, College Park, MD 20742, USA}

\author{F. Boschini}
\author{M. Zonno}
\affiliation{Department of Physics and Astronomy, University of British Columbia, Vancouver, BC V6T 1Z1, Canada}
\affiliation{Quantum Matter Institute, University of British Columbia, Vancouver, BC V6T 1Z4, Canada}
\author{M. Michiardi}
\affiliation{Department of Physics and Astronomy, University of British Columbia, Vancouver, BC V6T 1Z1, Canada}
\affiliation{Quantum Matter Institute, University of British Columbia, Vancouver, BC V6T 1Z4, Canada}
\affiliation{Max Planck Institute for Chemical Physics of Solids, N\"{o}thnitzer Straße 40, 01187 Dresden, Germany}

\author{E. Rotenberg}
\author{A. Bostwick}
\affiliation{Advanced Light Source, Lawrence Berkeley National Laboratory, Berkeley, CA 94720, USA}

\author{D. Graf}
\affiliation{National High Magnetic Field Laboratory, Florida State University, Tallahassee, FL 32306-4005, USA}

\author{B. J. Ramshaw}
\affiliation{Pulsed Field Facility, National High Magnetic Field Laboratory, Los Alamos National Laboratory, Los Alamos, NM 87545}
\affiliation{Laboratory of Atomic and Solid State Physics, Cornell University, 142 Sciences Drive, Ithaca, NY, 14853}

\author{A. Damascelli}
\affiliation{Department of Physics and Astronomy, University of British Columbia, Vancouver, BC V6T 1Z1, Canada}
\affiliation{Quantum Matter Institute, University of British Columbia, Vancouver, BC V6T 1Z4, Canada}
\affiliation{Canadian Institute for Advanced Research, Toronto, Canada M5G 1Z8}
\author{J. Paglione}
\affiliation{Maryland Quantum Materials Center, Department of Physics, University of Maryland, College Park, MD 20742, USA}
\affiliation{Canadian Institute for Advanced Research, Toronto, Canada M5G 1Z8}
\date{\today}

\begin{abstract}

Predictions of a topological electronic structure in the skutterudite TPn$_3$ family (T=transition metal, Pn=pnictogen) are investigated via magnetoresistance, quantum oscillations and angle-resolved photoemission experiments of RhSb$_3$, an unfilled skutterudite semimetal with low carrier density. Electronic band structure calculations and symmetry analysis of RhSb$_3$ indicate this material to be a zero-gap semimetal protected by symmetry with inverted valence/conduction bands that touch at the $\Gamma$ point close to the Fermi level. Transport experiments reveal an unsaturated linear magnetoresistance that approaches a factor of 200 at 60~T magnetic fields, and quantum oscillations observable up to 150~K that are consistent with a large Fermi velocity ($\sim 1.3\times 10^6$ m/s), high carrier mobility ($\sim 14$ m$^2$/Vs), and the existence of a small three dimensional hole pocket. A very small, sample-dependent effective mass falls to values as low as $0.018(2)$ of the bare electron mass and scales with Fermi wavevector. This, together with a non-zero Berry's phase and location of the Fermi level in the linear region of the valence band, suggests RhSb$_3$ as representative of a new class of toplogical semimeals with symmetry-enforced Fermi degeneracy at the high symmetry points.

\end{abstract}
\maketitle


Following the discovery of topological insulators \cite{TIReview1,TIReview2}, new classes of topological materials such as Dirac semimetals \cite{Young_2012}, Weyl semimetals  \cite{Wan_2011,Fang_2012}, nodal line and nodal chain semimetals~\cite{NodalChain,NodalLine}, have developed considerable interest. In Dirac systems, linearly dispersing valence and conduction bands touch at discrete points of four-fold degeneracy in the Brillouin zone (BZ), giving Dirac nodes protected against gap formation by crystal symmetry. Cd$_3$As$_2$ and Na$_3$Bi were the first theoretically predicted Dirac semimetal candidates \cite{A3Bi,Cd3As2-1}, later confirmed by experiments \cite{Borisenko_2014,Neupane_2014,Liu2014,Jeon_2014}. By breaking either inversion or time reversal symmetry, a Dirac semimetal can be tuned to a Weyl state where the nondegenerate linear touchings of the bulk bands come in pairs~\cite{Wan_2011,Fang_2012}, confirmed by recent experiments on materials such as the TaAs family (Type-I) \cite{TaAs-ARPES1,TaAs-ARPES2}. Recent studies have shown the existence of other exotic toplogical semimetal types, such as type-II Weyl semimetals like WTe$_2$ and MoTe$_2$~\cite{WTe2-Weyl,MoTe2-ARPES}, nodal-line/nodal-chain semimetals~\cite{CaAgAs_2016,IrF4_2016,ZrSiS_2016}, and triple fermions or beyond\cite{Triple-2,TopologicalQuantumChemisty,EightFold,TMDatabase,TMDatabasewebsite}. The topologically nontrivial band structure in such materials hosts unusual electronic states and exotic physical properties, including the so-called chiral anomaly phenomena and associated negative magnetoresistance (MR), non-local transport and the quantum anomalous Hall effect \cite{Liang_2014,Narayanan_2015,Wang2013a,Shekhar_2015,ZrTe5_Chiral,TaAs-Jia,TaAs-Chen,NbAs-Luo}. 
Interestingly, most Dirac and Weyl semimetals exhibit high mobilities and extremely large values of MR ~\cite{Liang_2014,Shekhar_2015,Narayanan_2015,TaAs-Jia,TaAs-Chen,NbAs-Luo,ZrSiS-MR}: carrier mobilities in Cd$_3$As$_2$, TaAs, and WTe$_2$ approach $100$ m$^2$/Vs, and MR scales of up to $\sim 5000$ are observed in 9 T fields. Extreme MR is believed to arise from the lifting of the high mobility of the Dirac/Weyl node protected by the (crystal or time reversal) symmetry or large fluctuations in the mobility due to the disorder effect of the Dirac quasiparticles~\cite{Liang_2014,Narayanan_2015}. Large MR also observed in some other semimetals such as LaSb and the NbSb$_2$ family~\cite{LaSb_Tafti,NbSb2-Wang,NbAs2_Shen,TaAs2_Luo} related to the topological band properties~\cite{LaBi_ARPES3,LaBi_ARPES2,LaBi_ARPES1,LaSb_ARPES1,NbSb2_Z2,NbSb2_HiddenWeyl}.

Although extensive studies have been reported, unique topological semimetal families are still rare. Seminal theoretical work by Singh {\it et al.} on the TPn$_3$ (T=transition metal, Pn=pnictogen) skutterudites that predated the topological revolution identified the possibility of an unusual quasi-linear band dispersion structure in the cubic unfilled skutterudite CoSb$_3$ \cite{CoSb3-PRL}. CoSb$_3$ is a well known thermoelectric small gap semiconductor \cite{Tang_2015,Singh_1994}, that, while topologically trivial, was predicted to be tunable through a topological quantum critical point via displacement of the Sb sublattices. Such a displacement was proposed to decrease gap size toward a critical point where the gap closes and massless (Dirac or Weyl) bands appear \cite{CoSb3-PRL,CoSb3-PRB}. With further displacement,  spin-orbit coupling would open a gap between inverted bands, realizing a non-trivial topological state. Similar behavior was also proposed in the theoretical compound IrBi$_3$, where band inversion between Ir-$d$ and Bi-$p$ orbtitals was predicted to occur \cite{IrBi3}. 
In this work, we report the first exploration of the intermediate member RhSb$_3$ using several experimental methods to identify a new class of topological semimetals with symmetry-enforced Fermi degeneracy at the high symmetry points, that is, enforced semimetals with Fermi degeneracy (ESFD)~\cite{TMDatabase,TMDatabasewebsite}, in this skutterudite series. Together with observations of very large MR and high carrier mobility values, quantum oscillations studies of the effective mass and non-trivial Berry's phase, as well as the angle-resolved photoemission spectroscopy (ARPES) experiments, are consistent with first-principle calculations of the electronic structure indicating a non-trivial topology is present in this family.

\begin{figure*}[tbp]
	\includegraphics[scale=0.65]{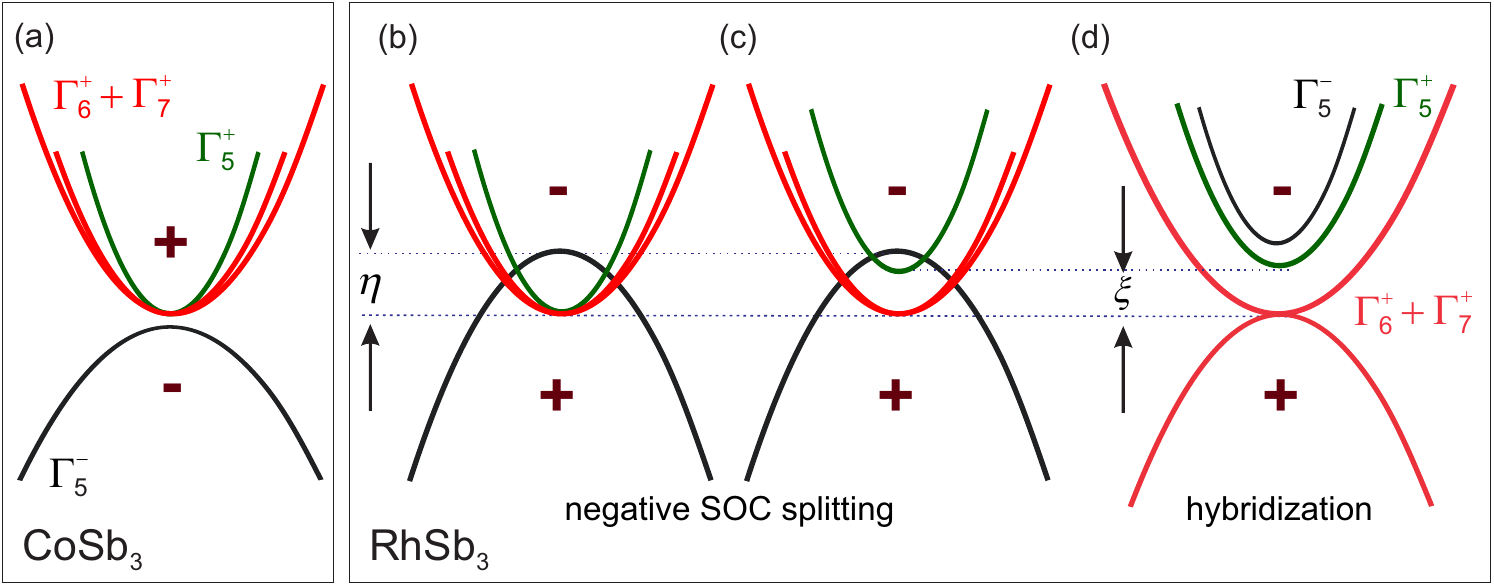}
	\caption{{\bf Band inversion and symmetry-enforced Fermi degeneracy of unfilled Skutterudite compound RhSb$_3$.} 
		(a)  Schematic of band ordering in CoSb$_3$ close to the Fermi level, showing the valence band with an irreducible representation $\Gamma_5^-$ and the conduction bands with irreducible representations $\Gamma_5^+$ and $\Gamma_6^++\Gamma_7^+$ touching at the $\Gamma$ point. (b) The stronger SOC in Rh atoms of RhSb$_3$ pushes the band $\Gamma_5^-$ up . The energy difference $\eta$ between $\Gamma_6^++\Gamma_7^+$ and $\Gamma_5^-$ indicating the strength of the band inversion, becomes negative. (c) The SOC in RhSb$_3$ also lift the degeneracy between $\Gamma_5^+$ and $\Gamma_6^++\Gamma_7^+$. (d) Finally, the overlap of the valence band $\Gamma_5^-$ and the conduction bands $\Gamma_5^+$ and $\Gamma_6^++\Gamma_7^+$ induces a strong hybridization between the p orbitals of Sb atoms and the d orbitals of Rh atoms which is evidenced by the calculated negative energy difference $\xi$ between $\Gamma_6^++\Gamma_7^+$ and $\Gamma_5^+$, giving the final band order of RhSb$_3$.}
\end{figure*}

\begin{figure*}[tbp]
	\includegraphics[scale=1]{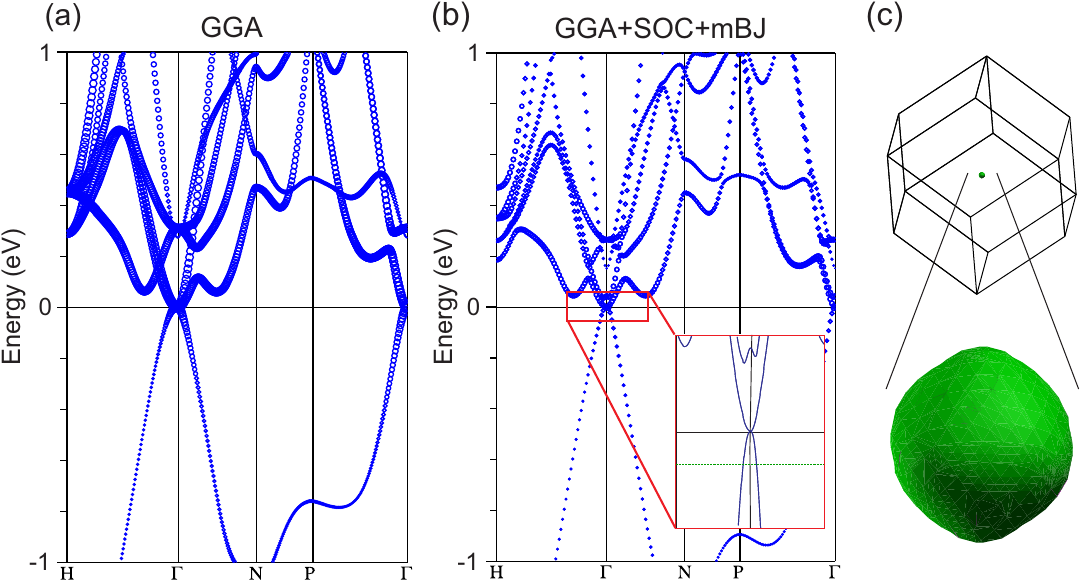}
	\caption{{\bf Calculated electronic structure of RhSb$_3$.} 
(a,b) The electronic structure of RhSb$_3$ was calculated from first-principles using two methods: (a) without spin-orbit coupling using PBE potential, and (b) with spin-orbit coupling and modified Becke-Johnson potential (see text). The radii of circles denoting band lines are drawn proportional Rh-$4d$ orbital weight. Inset in (b) An enlarged part of the band structure around the Fermi level. The dashed line indicates the position of the Fermi level for Sample D estimated by the carrier density from quantum oscillation and Hall resistivity. (c) The calculated Fermi surface of RhSb$_3$ is represented in the first Brillouin zone, and extended below and enlarged by a factor of 50 for clarity.}
\end{figure*}


The TSb$_3$ system with T=Co, Rh, and Ir, crystallizes in the symmorphic unfilled skutterudite structure (space group Im$\bar{3}$) with cubic Bravais lattice as shown in the inset of Fig. 1(a) and Fig. S1 in Supplementary Materials. Without SOC, for the space group Im$\bar{3}$, the $\Gamma$ point in the first Brillouin zone has a point group symmetry of $T_h$.  In CoSb$_3$, the highest valence band mainly derived from the $p$-orbitals of Sb is found to form an irreducible representation $\Gamma_1^-$ of the $T_h$ point group, while the three lowest conduction bands dominated by the $d-e_g$ orbitals of the (Co) transition metal atoms form an irreducible representation $\Gamma_4^+$. The CoSb$_3$ band structure is highly symmetric, with a single set of occupied and unoccupied bands touching at the high symmetry $\Gamma$ point. Turning on SOC lifts the degeneracy of these states and induces a gap of $\sim$50-120 meV between the valence band with an irreducible representation $\Gamma_5^-$ and the conduction bands with irreducible representations $\Gamma_5^+$ and $\Gamma_6^++\Gamma_7^+$, for the $m3$ double point group ~\cite{Singh_1994,CoSb3-PRL,Koga_2005,wzjcode}, as shown in Fig. 1(a)~\cite{CoSb3-PRL}. 

In RhSb$_3$, the stronger SOC in Rh atoms pushes the band $\Gamma_5^-$ up as shown in Fig. 1(b). The energy difference $\eta=E_{\Gamma_6^++\Gamma_7^+}-E_{\Gamma_5^-}$ indicating the strength of the band inversion, becomes negative. This implies the band inversion happens in RhSb$_3$ at the $\Gamma$ point. The SOC also lifts a degeneracy and opens a $\sim 20$ meV gap between the single band $\Gamma_5^+$ and the double-degenerated bands $\Gamma_6^++\Gamma_7^+$, in Fig. 1(c).  Further, the overlap of the valence band $\Gamma_5^-$ and the conduction bands $\Gamma_5^+$ and $\Gamma_6^++\Gamma_7^+$ induces a strong hybridization between the $p$ orbitals of Sb atoms and the $d$ orbitals of Rh atoms which is evidenced by the calculated negative $\xi=E_{\Gamma_6^++\Gamma_7^+}-E_{\Gamma_5^+}$ in Fig. 1(d). Correspondingly, in the highest valence band, the part far away from the $\Gamma$ point is mainly contributed from the $p$-orbitals of Sb atoms. However, the weight of Rh-$4d$ orbitals (indicated by the radii of circles in the band line of Fig. 2(a,b)) increases when approaching the Fermi level and  becomes dominant. This clearly indicates a band inversion between Rh$-d$ and Sb-$p$ orbitals. The highest valence band and the lowest conduction band form the irreducible representations $\Gamma_6^++\Gamma_7^+$, and the band $\Gamma_5^-$ sits above all these bands. However, there is no influence on the double-degenerated bands $\Gamma_6^++\Gamma_7^+$ at $\Gamma$ because the double degeneracy is enforced by the $m3$ double point group. Therefore the Fermi energy of RhSb$_3$ is pinned to the doublet $\Gamma_6^++\Gamma_7^+$, and is protected by symmorphic crystalline symmetry and time reversal symmetry~\cite{CoSb3-PRB}. As shown in the band structure in Figs.~2(b), a Dirac-like quasi-linear dispersion persists over a wide energy range, with a more parabolic dispersion appearing in a very small energy window of $\sim 20$~meV close to the Fermi level as shown in Fig. 2(b). The resultant calculated Fermi surface is an extremely small isotropic hole pocket centered at the $\Gamma$ point (Fig. 2(c)). Together, the band inversion and degeneracy near the Fermi level point to RhSb$_3$ as a new type of ESFD. The ESFD is easily tuned to a topological insulator by symmetry breaking such as a uniaxial strain in the $z$ direction. In other words, an infinitesimal symmetry breaking perturbation can lead it to a topological insulating phase, which is confirmed by the Fu-Kane $Z_2$ invariants ($\nu_0=1, \nu_1=\nu_2=\nu_3=0$) in the calculation with uniaxial strain (shown in Fig. S2(b) in Supplementary Materials).

\begin{figure*}[tbp]
	\includegraphics[scale=1]{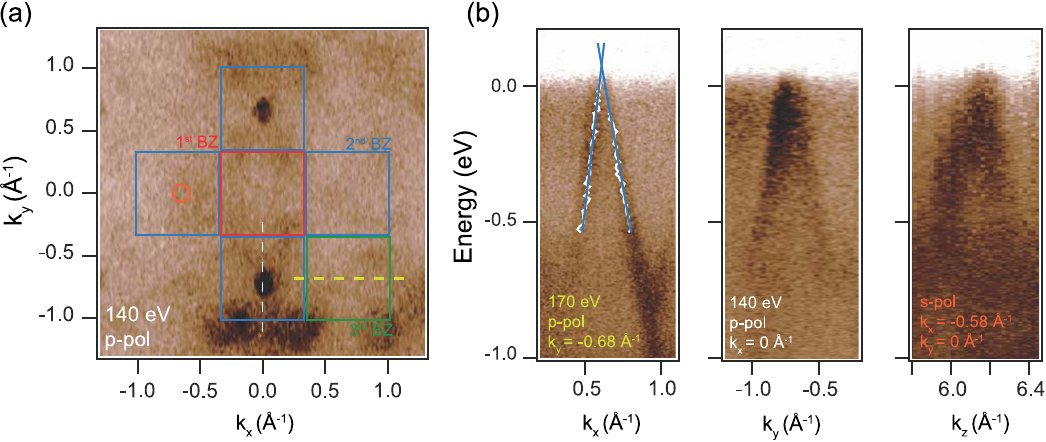}
	\caption{ {\bf Linear dispersion of RhSb$_3$ band structure from photon-energy dependent ARPES at $T$=80 K.} 
(a) Fermi surface at 140 eV, $p$-polarization. The red square shows the first surface-projected Brillouin zone for the (100) surface, while blue and green squares are the second and third zones (see more details in the Supplementary Materials). 
(b) ARPES band mapping along $k_x$, $k_y$ and $k_z$ directions. Left panel: band mapping along $k_y=-0.68$~\AA$^{-1}$ line in the third BZ (yellow dashed line in (a), 170 eV ($k_z\sim5\times2\pi/c$), $p$-polarization). White solid curves show the peak position of the momentum distribution curves from a double-Gaussian fit, while blue solid lines are the result of a linear fit of the band dispersion. Central panel: band mapping along $k_x=0$~\AA$^{-1}$ line in the second BZ (white dashed line in (a), 140 eV ($k_z\sim4.5\times2\pi/c$), $p$-polarization). Right panel: $k_z$ dependence, assuming 10 eV inner potential, of the hole pocket in the second BZ at $(k_x, k_y)=(-0.57$~\AA$^{-1}, 0$~\AA$^{-1})$; orange circles in (a), $s$-polarization).}
\end{figure*}


\begin{figure*}[tbp]
	\includegraphics[scale=0.6]{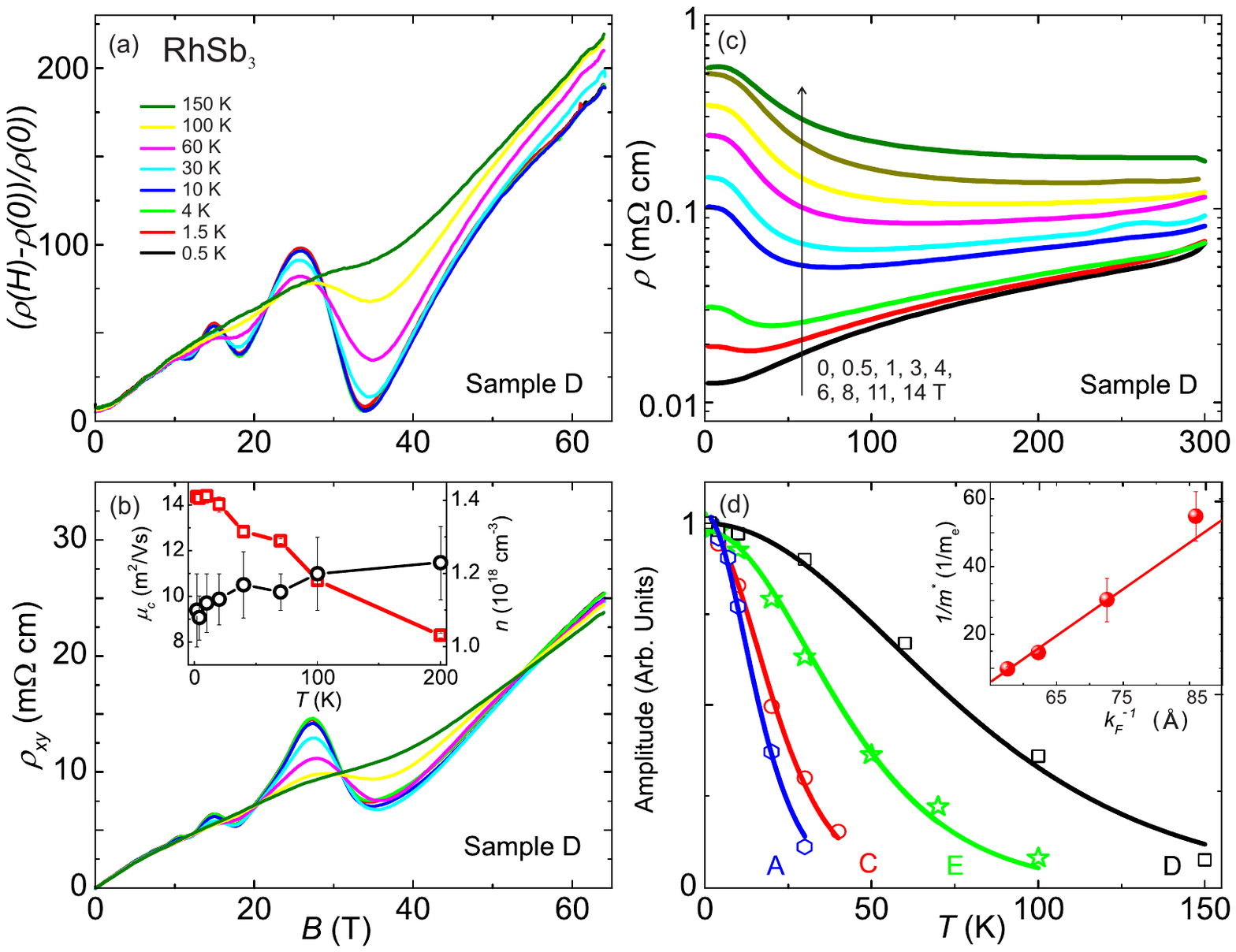}
	\caption{{\bf Large magnetoresistance, high carrier mobility and quantum oscillations of RhSb$_3$ single crystals.} 
(a) Magnetoresistance of RhSb$_3$ sample D for field orientation transverse to current direction, showing very large non-saturating enhancement up to 65~T pulsed field and prominent Shubnikov-de Haas oscillations up to temperatures of 150~K.
(b) Hall resistivity $\rho_{xy}$ of sample D (same temperature values as panel (a)), with carrier mobility and density extracted from fitting of data shown in inset. 
(c) Temperature dependence of the resistivity of sample D at different magnetic fields. 
(d) Amplitudes of Shubnikov-de Haas oscillations of the Fermi pocket as a function of temperature $T$, for four samples with varying carrier densities. The solid lines are fits to the Lifshitz-Kosevich formula for samples A, C, D and E, which exhibit effective mass ratios $m^*/m_e$ of 0.102(8), 0.066(7), 0.018(2) and 0.033(4), respectively, corresponding to oscillation frequencies of 71, 60, 33 and 43~T, respectively. The inset shows the linear relation of $\frac{1}{m^*}$ versus $k_F^{-1}=\sqrt{\frac{\pi}{F}}$ for these samples (solid symbols), revealing the Dirac-like dispersion (solid line fit) for RhSb$_3$.}
\end{figure*}


To evaluate and confirm the non-trivial topology of these bands, we performed both ARPES and quantum oscillations experiments. 
Photon energy-dependent ARPES measurements confirm the presence of a linearly-dispersive three-dimensional hole pocket around $\Gamma$.  The sample was cleaved to expose the (001) surface. Assuming a 10~eV inner potential, at 140~eV photon energy the momentum probed along the $c$-axis corresponds to $k_z\sim 4.5\times2\pi/c$ ($\pi/c=\Gamma-H$). As a consequence of the body-centered cubic structure, the $\Gamma$ points from adjacent surface-projected BZs are characterized by a $\pi/c$ offset in $k_z$. For this reason, hole-like Fermi surface pockets can be observed in Fig. 3(a) only around the $\Gamma$ points of the second BZ. Furthermore, due to matrix element effects, when using linear horizontal polarization (p), the second BZ hole pockets are detected only along the $k_x=0$~\AA$^{-1}$ line while, with linear vertical polarization (s), those same pockets are observed exclusively along the $k_y=0$~\AA$^{-1}$ line (See Fig. S3 of Supplementary Materials). 

The resultant ARPES band mapping along $k_x$, $k_y$ and $k_z$ directions is presented in Fig. 3(b). As a consequence of matrix elements effects and to maximize the photoemission intensity from the hole pocket, these three cuts have been acquired in different regions of momentum space using different photon energies and incident polarizations. The left-hand side and central panels of Fig. 3(b) present the band mapping along the $k_x$ and $k_y$ directions (yellow and white dashed lines in (a), respectively). By fitting the momentum distribution curves in the left panel, we estimate a Fermi velocity of $(0.6 \pm 0.2)\times10^6$ m/s ($4\pm1.2$~eV \AA). In order to verify the three-dimensional nature of the detected pocket, photon-energy scans were performed to probe the $k_z$ dependence. As shown in Fig. 3(b) right panel, the photoemission intensity at $(k_x, k_y)=(-0.57$~\AA$^{-1}, 0$~\AA$^{-1})$ (slightly off the $\Gamma$ point) changes as a function of energy, qualitatively defining a nearly isotropic three-dimensional cone-like dispersion for the hole pocket. This is in good agreement with band structure calculations, and for comparison is in contrast to observations of Cd$_2$As$_3$, where the Fermi velocity is markedly anisotropic~\cite{Liu2014}.
Furthermore, the field angular dependence of the frequency of quantum oscillations in RhSb$_3$ crystal (shown in Fig. S4 of Supplementary Materials) exhibits very little variation with field angle orientation, confirming the case for a nearly spherical bulk three-dimensional Fermi pocket.

Magnetotransport and Shubnikov-de Haas (SdH) quantum oscillations experiments were performed on four representative samples. 
Magnetic fields applied along the principle axis in a transverse configuration significantly enhance the resistivity of RhSb$_3$. The MR ($=(\rho_{xx}(B)-\rho_{xx}(0))/\rho_{xx}(0)$) is linear in the high field region as evidenced by the field derivative of MR at different temperature respectively for Sample D shown in Fig. S8 in the Supplementary Materials, and does not show any sign of saturation up to a 64 T pulsed field (Fig. 4(a)) but exhibits quantum oscillations above  10 T (also observed in Hall resistivity, Fig. 4(b)), and continues to increase above the quantum limit of $\sim 35$ T. The MR ratio is close to the reported value ($10^3 \sim 10^6$) observed in both the Dirac material Cd$_3$As$_2$ and the TaAs family of Weyl materials~\cite{Liang_2014,Shekhar_2015,Narayanan_2015,TaAs-Jia,TaAs-Chen,NbAs-Luo,ZrSiS-MR}
As shown in Fig.~4, the MR has a strong temperature dependence, driving a semi-metallic character in zero field toward a semiconductor-like behavior with a saturating resistivity at the lowest temperatures.
While the absolute MR amplitude varies among different samples and is dependent on the residual resistivity ratio of the crystals, all crystals show a crossover from semimetallic character to semiconductor-like behavior driven by magnetic field (see Figs. S5 and Fig. S6 in Supplementary Materials for more details). 

Fig. 4(b) presents the Hall resistivity $\rho_{xy}$ of RhSb$_3$ as a function of field at different temperatures. In fields below 6 T, $\rho_{xy}$ is linear and nearly temperature-independent, while showing large quantum oscillations at higher fields that degrade with increasing temperature. The carrier density $n$ of sample D is calculated by linear fits of the low-field $\rho_{xy}$, yielding a value $n\sim 1.1\times10^{18}$ cm$^{-3}$ at low temperature and only $\sim 10\%$ variation upon increasing temperature as shown in the inset of Fig. 4(b). This is consistent with the zero-gap band structure depicted in Fig.~2. From the carrier density $n$ and the zero-field resistivity $\rho_{xx}=18~ \mu\Omega$cm, the classical mobility $\mu_c=1/\rho_{xx}ne$ is calculated to approach 14 m$^2$/Vs at 1.8 K, nearly that of graphene, Cd$_3$As$_2$ and TaAs/TaP, which fall in the range 1-100 m$^2$/Vs.

In addition to MR variations, SdH oscillations measurements performed on four different samples indicate a tunability of the oscillation frequency, and hence the carrier density and corresponding chemical potential variation.
Quantum oscillations in conductivity are given by
\begin{equation}
\Delta\sigma_{xx}(T,B)=A(T,B)\cos\{2\pi[(F/B)-1/2+\beta+\delta]\},
\end{equation}
with the non-oscillating amplitude $A(T,B)=e^{-2\pi^2k_BT_D/\hbar\omega_c}\frac{2\pi^2k_BT/\hbar\omega_c}{\sinh(2\pi^2k_BT/\hbar\omega_c)}$, Dingle temperature $T_D$, cyclotron frequency $\omega_c=eB/m^*$ with effective mass $m^*$, phase factor $-1/2+\beta+\delta$ and SdH frequency $F$ of the oscillation corresponding to the cross-section of Fermi surface defined by the cyclotron orbits~\cite{OscillationBook}. 
For a trivial parabolic dispersion, $\beta$=0 and therefore the Berry's phase is zero, while $\beta$=1/2 for a Dirac dispersion giving a Berry's phase of $\pi$. $\delta$ is a phase shift resulting from the curvature of the Fermi surface in the third direction, taking a value of $\delta=0$ ($\pm\frac{1}{8}$) for a two-dimensional (three dimensional) Fermi surface.\cite{phase-shift1,phase-shift2,phase-shift3,phase-shift4} The frequency $F$ is given by the Onsanger relation $F=\frac{\hbar\pi k_F^2}{2\pi e}$ with $k_F$ being the Fermi wave vector in the spherical Fermi surface approximation.

\begin{figure}[tbp]
	\includegraphics[scale=0.4]{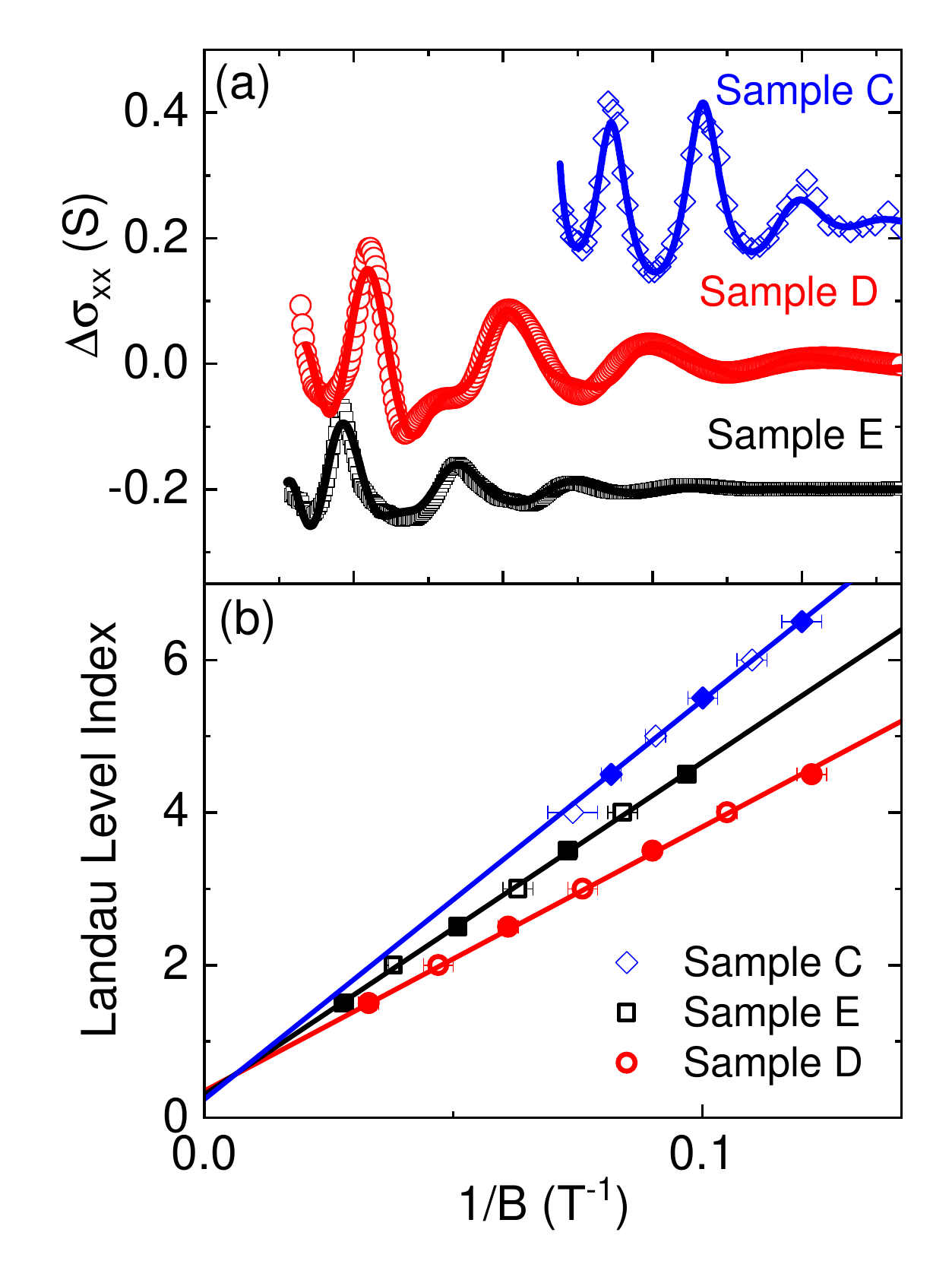}
	\caption{{\bf Berry phase analysis of Shubnikov-de Haas oscillations in RhSb$_3$.}
(a) Quantum oscillations of background subtracted conductivity $\Delta\sigma_{xx}$ as a function of inverse fields up to 64 T for samples D and E. The discrete symbols are the experimental data, while the lines are fits using a global fitting routine as described in the text. (b) Landau level (LL) index plot of oscillations of three samples, with integer levels assigned to the minima of $\Delta\sigma_{xx}$ and maxima assigned as half-integer indices. Solid lines are linear fits to the data, extrapolating to infinite field intercepts that correspond to finite Berry's phase values of $0.92(1)\pi$, $1.02(1)\pi$ and $0.98(1)\pi$ for samples C, D and E, respectively (see text).}
\end{figure}

As evident in Fig.~4, the SdH oscillations of sample D have a very small frequency, which is 33~T as extracted from FFT analysis of the background-subtracted resistivity (see SM Fig.~S6 for FFT spectra of all samples). This value corresponds to wave vector $k_F=0.04(5)~ \AA^{-1}$ and a very small carrier density $n_{SdH}=0.8(1)\times10^{18}$~cm$^{-3}$, which is consistent with Hall resistivity results. 
The other three crystals also exhibit SdH oscillations (see Fig. S7 in Supplementary Materials) with varying amplitudes. FFT analysis yield a relatively large variation in frequencies: 71~T, 60~T and 43~T for samples A, C and E, respectively. 
Moreover, oscillations observed in Sample D appear to decay very slowly with increasing temperature, still being observable at 150 K. 
Interestingly, the amplitude of oscillations appears to be suppressed at different temperatures for each sample. While oscillations are still evident at 150~K for sample D (Fig.~4), they disappear near 30~K, 50~K and 100~K for samples A, C and E, respectively.
This suggests that samples with smaller frequency (or carrier density) have charge carriers with systematically smaller effective masses. 
The decaying of the amplitude of SdH oscillation with temperature is described by the Lifshitz-Kosevich (LK) formula $A(T,B)$ given by Eqn.(2). 
As shown in Fig.~4(d), tracking the FFT amplitudes of all four samples as a function of temperature indeed shows a systematic trend. Fitting to the LK formula gives effective mass ratios $m^*/m_e$ of 
0.102(8), 0.066(7), 0.033(4) and 0.018(2) for samples A, C, E and D, respectively, in descending mass order.

The effective mass in RhSb$_3$ is extremely light, being rivaled only by InSb, which has an effective mass $\sim 0.014 m_e$ that yields observable oscillations up to 175 K~\cite{InSb1,InSb2}.  The effective mass in RhSb$_3$ is also in fact comparable to those observed in typical Dirac ($0.023m_e$ in Cd$_3$As$_2$ \cite{Liang_2014,Narayanan_2015}) and Weyl ($0.033m_e$ in NbAs \cite{NbAs-Luo}, $0.076m_e$ in NbP \cite{Shekhar_2015} and $0.15m_e$ in TaAs \cite{Huang2015}) semimetals. 
However, in RhSb$_3$ the systematic variation of $m^*$ with $F$ is indicative of a departure from the expectations for a parabolic dispersion. The effective mass, defined as $m^*=\hbar k_F/v_F$ should be constant for a parabolic band with different carrier densities since the Fermi velocity $v_F=\frac{1}{\hbar}\frac{\partial\varepsilon}{\partial k}|_{k_F}\propto k_F$ with $\varepsilon$ being the band energy. However this is not true for Dirac fermions with a linear dispersion, where $v_F$ is a constant. For instance, in graphene it is expected that $m^*$ is proportional to the square root of carrier density since $k_F=(\pi n_s)^{1/2}$, which was indeed confirmed by several experiments \cite{Graphene_RMP,Graphene_Novoselov_2005} Similar phenomenon has also been shown in Cd$_3$As$_2$ \cite{Narayanan_2015}. 
In RhSb$_3$, the relation between $m^*$ and $k_F$ follows the equation above very well, as shown for all four samples in Fig. 4(d), except for an intercept which is possibly due to the breakdown of the semiclassical transport theory close to the quantum limit. This giving a constant $v_F = 1.3\times10^6$~m/s that is close to the ARPES value and also to that of Cd$_3$As$_2$ ($\sim 4\times10^6$ m/s) and NdP ($\sim 4.8\times10^5$ m/s).

The observation of a constant Fermi velocity with varying carrier concentration in RhSb$_3$ directly confirms that the bulk carriers have a linear dispersion very close to the degeneracy point, and the extreme values of $m^*$ and $v_F$ are considered to be responsible for the high mobility values.
The topological nature of this band structure can be further explored by analysis of the phase factor of the quantum oscillations as indicated in Eq. (1). 
Experimentally, this phase shift in the semi-classical regime can be obtained from
an analysis of the relation between the Landau level (LL) index $N$ and energy, widely represented in a so-called fan diagram of integer $N$ plotted as a function of inverse field $1/B$. The slope of such a plot is dependent on the oscillation frequency $F$ and the y-axis intercept yields the Berry's phase $\beta$, in units of $2\pi$. For a topologically trivial material with parabolic dispersion, an intercept of zero is expected, while that with a chiral Dirac dispersion would possess finite Berry curvature and therefore a Berry's phase of value $\pi$.

Fig. 5(a) presents the oscillatory component of the conductivity $\Delta\sigma_{xx}$ of samples C, D and E as a function of $1/B$. Assigning the minima and maxima of $\Delta\sigma_{xx}$ to integer ($N$) and half-integer ($N+1/2$) indices (for the oscillation in high field part, the position of minima is defined as an average of two sub-minima due to the spin splitting), respectively, yields the LL fan diagram as shown in Fig. 5(b), demonstrating the variation in slopes that correspond to the different carrier densities of each sample. 
A linear extrapolation of $N$ versus $1/B$ to the infinite field limit for each sample yields finite intercepts of $0.352(4), 0.381(2)$ and $0.373(6)$ for samples C, D and E, respectively.
To properly extract the Berry's phase values, one must consider the phase shift due to geometry. 
Taking the $\delta=-\frac{1}{8}$ phase shift due to the spherical geometry of the Fermi surface (the hole pocket) of RhSb$_3$ into account, the measured intercept values are very close to the adjusted value $3/8$ corresponding to the geometry-adjusted phase shift of $1/2+\delta$, yielding a phase factor $\beta=\gamma+\frac{1}{8}$ for a Berry's phase of $0.92(1)\pi$, $1.02(1)\pi$ and $0.98(1)\pi$ for samples C, D and E, respectively. 
Because the linear extrapolation of the LL indices is a simplistic fit of the LL spectrum, we have implemented a global fitting routine to further confirm the nontrivial Berry's phase in this study.
Using Eq. 1, we fit $\Delta\sigma_{xx}$ of samples C, D and E using the "Bumps" code (see Methods section) \cite{Bumps}, incorporating a Zeeman splitting term $F\pm\delta_F$ to capture the observable splitting to yield well-converging fits (shown in SI materials Figs. S9 and S10) to the experimental data as shown in Fig. 5(a). The Berry's phases extracted from this global fitting procedure are $0.87(1)\pi$, $0.96(1)\pi$ and $0.81(1)\pi$ for samples C, D and E, respectively, consistent with the LL index extrapolations and therefore unequivocally implying the presence of a non-trivial topology and the existence of Dirac quasiparticles in the bulk of RhSb$_3$. 
While the energy resolution is not enough in both experiment (ARPES) and theoretical calculation to resolve the exact nature of the dispersion at the degeneracy point (i.e. linear or quadratic), the lack of any observable curvature in the fan diagram suggests that the dispersion is very linear and Dirac-like~\cite{ando}, prompting further theoretical consideration of the exact nature of topology in this new class of zero-gap topological semimetals.

\section{Methods}

Single crystals of RhSb$_3$ were synthesized by a high-temperature self-flux method. X-ray diffraction data were taken at room temperature with Cu K$_{\alpha}$ ($\lambda=0.15418$ nm) radiation in a powder diffractometer. Electrical transport measurements up to 14 T were conducted on polished samples in a commercial cryostat applying the conventional four-wire contact method. High field MR and Shubnikov-de Haas (SdH) oscillation were measured at the National High Magnetic Field Laboratory (NHMFL) Tallahassee DC Field Facility up to 35 T, using a one-axis rotator in a He-3 cryostat with temperature range 0.35 K $\sim$ 70 K, and also at the NHMFL Los Alamos pulsed field facility up to 65 T. All measurements were performed with electrical current flowing along the [100] direction. Magnetic field direction was always kept perpendicular to current, but rotated from perpendicular to (100) face ($\theta=0^o$) to direction along (100) face ($\theta=90^o$).
The longitudinal conductivity $\sigma_{xx}$ is derived through tensor conversion from the longitudinal resistivity $\rho_{xx}$ and the Hall resistivity $\rho_{xy}$ by $\sigma_{xx}=\frac{\rho_{xx}}{\rho_{xx}^2+\rho_{xy}^2}$.

The global fitting and uncertainty analysis for the quantum oscillation in the longitudinal conductivity $\sigma_{xx}$ for BaAl$_4$ single crystals was performed using a Broyden-Fletcher$-$Goldfarb-Shanno (BFGS) optimizer and a Markov chain Monte Carlo (MCMC) method to compute the joint distribution of parameter probabilities and to find the global minimum, implemented in the ``Bumps'' code.\cite{Bumps} Bumps is a set of free and public routines for complicated curve fitting, uncertainty analysis and correlation analysis from a Bayesian perspective. Bumps provides uncertainty analysis which explores all viable minima using the MCMC method and finds confidence intervals on the parameters based on uncertainty from experimental errors. For full uncertainty analysis, bumps further uses a random walk to explore the viable parameter space near the minimum, showing pair-wise correlations between the different parameter values. The 2D correlation plots indicate the correlation relationship between multiple parameters in the fitting function.

ARPES measurements were performed at beamline 7.0.2.1 of the Advanced Light Source (ALS) at Lawrence Berkeley National Laboratory. The sample was cleaved and kept at 80 K at a base pressure of $5\times 10^{-11}$ Torr. Photoemitted electrons were detected with a Scienta R4000 analyser, for photon energy ranging from 130 to 170 eV. The energy and the angular resolution were better than 50 meV and 0.2$^o$, respectively. 

First principles electronic structure calculations were performed within the full-potential linearized augmented plane wave (LAPW) method \cite{wien2k1} implemented in WIEN2k package, without or with spin-orbit coupling.\cite{wien2k2} The general gradient approximation (GGA) of Perdew \textit{et al}.,\cite{gga} was used for exchange-correlation potential. The LAPW sphere radius were set to 2.5 Bohr for all atoms. The converged basis corresponding to $R_{min}k_{max}=7$ with additional local orbital were used, where $R_{min}$ is the minimum LAPW sphere radius and $k_{max}$ is the plane wave cutoff. Lattice parameters were obtained from refinement of the x-ray diffraction pattern obtained from powdered single crystals. In the calculation with spin-orbit coupling, the modified Becke-Johnson (mBJ) exchange potential was combined with the GGA correlation.\cite{mBJ} It can effectively mimic the behavior of orbital dependent potential around the band gap, so is expected to obtain the accurate position of states near the band edge and the band order, which are the keys to determine the band inversion and the band topology. For high-symmetry points, the eigenvalues (irreducible represenations) of all symmetry operators have been computed from the wave functions for each band using the open-source code vasp2trace~\cite{wzjcode}, referring to the character tables in the Bilbao Crystallographic Server (BCS)~\cite{server,crystallographicserver}.

\section{acknowledgments}
The authors acknowledge useful discussions with A. Bernevig, Z.J. Wang, and J. Sau. 
This work was supported by the Gordon and Betty Moore Foundation's EPiQS Initiative through Grant GBMF9071. A portion of this work was performed at the National High Magnetic Field Laboratory, which is supported by National Science Foundation Cooperative Agreement No. DMR-1157490 and the State of Florida. This research was undertaken thanks in part to funding from the Max Planck–UBC Centre for Quantum Materials and the Canada First Research Excellence Fund, Quantum Materials and Future Technologies Program. The work at UBC was supported by the Killam, Alfred P. Sloan, and Natural Sciences and Engineering Research Council of Canada’s (NSERC’s) Steacie Memorial Fellowships (A.D.); the Alexander von Humboldt Fellowship (A.D.); the Canada Research Chairs Program (A.D.); and the NSERC, Canada Foundation for Innovation (CFI), and CIFAR Quantum Materials.

\bibliographystyle{naturemag}
\bibliography{TopologicalMaterials}

\begin{thebibliography}{10}
\expandafter\ifx\csname url\endcsname\relax
  \def\url#1{\texttt{#1}}\fi
\expandafter\ifx\csname urlprefix\endcsname\relax\def\urlprefix{URL }\fi
\providecommand{\bibinfo}[2]{#2}
\providecommand{\eprint}[2][]{\url{#2}}

\bibitem{TIReview1}
\bibinfo{author}{Hasan, M.~Z.} \& \bibinfo{author}{Kane, C.~L.}
\newblock \bibinfo{title}{Colloquium : Topological insulators}.
\newblock \emph{\bibinfo{journal}{Reviews of Modern Physics}}
  \textbf{\bibinfo{volume}{82}}, \bibinfo{pages}{3045--3067}
  (\bibinfo{year}{2010}).
\newblock \urlprefix\url{http://dx.doi.org/10.1103/RevModPhys.82.3045}.

\bibitem{TIReview2}
\bibinfo{author}{Qi, X.-L.} \& \bibinfo{author}{Zhang, S.-C.}
\newblock \bibinfo{title}{Topological insulators and superconductors}.
\newblock \emph{\bibinfo{journal}{Reviews of Modern Physics}}
  \textbf{\bibinfo{volume}{83}}, \bibinfo{pages}{1057--1110}
  (\bibinfo{year}{2011}).
\newblock \urlprefix\url{http://dx.doi.org/10.1103/RevModPhys.83.1057}.

\bibitem{Young_2012}
\bibinfo{author}{Young, S.~M.} \emph{et~al.}
\newblock \bibinfo{title}{Dirac semimetal in three dimensions}.
\newblock \emph{\bibinfo{journal}{Phys. Rev. Lett.}}
  \textbf{\bibinfo{volume}{108}}, \bibinfo{pages}{140405}
  (\bibinfo{year}{2012}).
\newblock \urlprefix\url{http://dx.doi.org/10.1103/PhysRevLett.108.140405}.

\bibitem{Wan_2011}
\bibinfo{author}{Wan, X.}, \bibinfo{author}{Turner, A.~M.},
  \bibinfo{author}{Vishwanath, A.} \& \bibinfo{author}{Savrasov, S.~Y.}
\newblock \bibinfo{title}{Topological semimetal and fermi-arc surface states in
  the electronic structure of pyrochlore iridates}.
\newblock \emph{\bibinfo{journal}{Phys. Rev. B}} \textbf{\bibinfo{volume}{83}},
  \bibinfo{pages}{205101} (\bibinfo{year}{2011}).
\newblock \urlprefix\url{http://dx.doi.org/10.1103/PhysRevB.83.205101}.

\bibitem{Fang_2012}
\bibinfo{author}{Fang, C.}, \bibinfo{author}{Gilbert, M.~J.},
  \bibinfo{author}{Dai, X.} \& \bibinfo{author}{Bernevig, B.~A.}
\newblock \bibinfo{title}{Multi-weyl topological semimetals stabilized by point
  group symmetry}.
\newblock \emph{\bibinfo{journal}{Phys. Rev. Lett.}}
  \textbf{\bibinfo{volume}{108}}, \bibinfo{pages}{266802}
  (\bibinfo{year}{2012}).
\newblock \urlprefix\url{http://dx.doi.org/10.1103/PhysRevLett.108.266802}.

\bibitem{NodalChain}
\bibinfo{author}{Bzdu{\v{s}}ek, T.}, \bibinfo{author}{Wu, Q.},
  \bibinfo{author}{Rüegg, A.}, \bibinfo{author}{Sigrist, M.} \&
  \bibinfo{author}{Soluyanov, A.~A.}
\newblock \bibinfo{title}{Nodal-chain metals}.
\newblock \emph{\bibinfo{journal}{Nature}} \textbf{\bibinfo{volume}{538}},
  \bibinfo{pages}{75--78} (\bibinfo{year}{2016}).
\newblock \urlprefix\url{http://dx.doi.org/10.1038/nature19099}.

\bibitem{NodalLine}
\bibinfo{author}{Burkov, A.~A.}, \bibinfo{author}{Hook, M.~D.} \&
  \bibinfo{author}{Balents, L.}
\newblock \bibinfo{title}{Topological nodal semimetals}.
\newblock \emph{\bibinfo{journal}{Phys. Rev. B}} \textbf{\bibinfo{volume}{84}},
  \bibinfo{pages}{235126} (\bibinfo{year}{2011}).
\newblock \urlprefix\url{http://dx.doi.org/10.1103/PhysRevB.84.235126}.

\bibitem{A3Bi}
\bibinfo{author}{Wang, Z.} \emph{et~al.}
\newblock \bibinfo{title}{Dirac semimetal and topological phase transitions in
  a 3~{B}i ( a = na , k, rb)}.
\newblock \emph{\bibinfo{journal}{Phys. Rev. B}} \textbf{\bibinfo{volume}{85}},
  \bibinfo{pages}{195320} (\bibinfo{year}{2012}).
\newblock \urlprefix\url{http://dx.doi.org/10.1103/PhysRevB.85.195320}.

\bibitem{Cd3As2-1}
\bibinfo{author}{Wang, Z.}, \bibinfo{author}{Weng, H.}, \bibinfo{author}{Wu,
  Q.}, \bibinfo{author}{Dai, X.} \& \bibinfo{author}{Fang, Z.}
\newblock \bibinfo{title}{Three-dimensional dirac semimetal and quantum
  transport in cd 3~{A}s 2}.
\newblock \emph{\bibinfo{journal}{Phys. Rev. B}} \textbf{\bibinfo{volume}{88}},
  \bibinfo{pages}{125427} (\bibinfo{year}{2013}).
\newblock \urlprefix\url{http://dx.doi.org/10.1103/PhysRevB.88.125427}.

\bibitem{Borisenko_2014}
\bibinfo{author}{Borisenko, S.} \emph{et~al.}
\newblock \bibinfo{title}{Experimental realization of a three-dimensional dirac
  semimetal}.
\newblock \emph{\bibinfo{journal}{Phys. Rev. Lett.}}
  \textbf{\bibinfo{volume}{113}}, \bibinfo{pages}{027603}
  (\bibinfo{year}{2014}).
\newblock \urlprefix\url{http://dx.doi.org/10.1103/PhysRevLett.113.027603}.

\bibitem{Neupane_2014}
\bibinfo{author}{Neupane, M.} \emph{et~al.}
\newblock \bibinfo{title}{Observation of a three-dimensional topological dirac
  semimetal phase in high-mobility cd3{A}s2}.
\newblock \emph{\bibinfo{journal}{Nature Communications}}
  \textbf{\bibinfo{volume}{5}} (\bibinfo{year}{2014}).
\newblock \urlprefix\url{http://dx.doi.org/10.1038/ncomms4786}.

\bibitem{Liu2014}
\bibinfo{author}{Liu, Z.~K.} \emph{et~al.}
\newblock \bibinfo{title}{A stable three-dimensional topological dirac
  semimetal cd3{A}s2}.
\newblock \emph{\bibinfo{journal}{Nature Materials}}
  \textbf{\bibinfo{volume}{13}}, \bibinfo{pages}{677--681}
  (\bibinfo{year}{2014}).
\newblock \urlprefix\url{http://dx.doi.org/10.1038/nmat3990}.

\bibitem{Jeon_2014}
\bibinfo{author}{Jeon, S.} \emph{et~al.}
\newblock \bibinfo{title}{Landau quantization and quasiparticle interference in
  the three-dimensional dirac~semimetal cd3{A}s2}.
\newblock \emph{\bibinfo{journal}{Nature Materials}}
  \textbf{\bibinfo{volume}{13}}, \bibinfo{pages}{851--856}
  (\bibinfo{year}{2014}).
\newblock \urlprefix\url{http://dx.doi.org/10.1038/nmat4023}.

\bibitem{TaAs-ARPES1}
\bibinfo{author}{Lv, B.~Q.} \emph{et~al.}
\newblock \bibinfo{title}{Experimental discovery of weyl semimetal {TaAs}}.
\newblock \emph{\bibinfo{journal}{Phys. Rev. X}} \textbf{\bibinfo{volume}{5}},
  \bibinfo{pages}{031013} (\bibinfo{year}{2015}).
\newblock \urlprefix\url{http://dx.doi.org/10.1103/PhysRevX.5.031013}.

\bibitem{TaAs-ARPES2}
\bibinfo{author}{Xu, S.-Y.} \emph{et~al.}
\newblock \bibinfo{title}{Discovery of a weyl fermion semimetal and topological
  fermi arcs}.
\newblock \emph{\bibinfo{journal}{Science}} \textbf{\bibinfo{volume}{349}},
  \bibinfo{pages}{613--617} (\bibinfo{year}{2015}).
\newblock \urlprefix\url{http://dx.doi.org/10.1126/science.aaa9297}.

\bibitem{WTe2-Weyl}
\bibinfo{author}{Soluyanov, A.~A.} \emph{et~al.}
\newblock \bibinfo{title}{Type-{II} weyl semimetals}.
\newblock \emph{\bibinfo{journal}{Nature}} \textbf{\bibinfo{volume}{527}},
  \bibinfo{pages}{495--498} (\bibinfo{year}{2015}).
\newblock \urlprefix\url{http://dx.doi.org/10.1038/nature15768}.

\bibitem{MoTe2-ARPES}
\bibinfo{author}{Huang, L.} \emph{et~al.}
\newblock \bibinfo{title}{Spectroscopic evidence for a type {II} weyl
  semimetallic state in {MoTe}2}.
\newblock \emph{\bibinfo{journal}{Nature Materials}}
  \textbf{\bibinfo{volume}{15}}, \bibinfo{pages}{1155--1160}
  (\bibinfo{year}{2016}).
\newblock \urlprefix\url{http://dx.doi.org/10.1038/nmat4685}.

\bibitem{CaAgAs_2016}
\bibinfo{author}{Yamakage, A.}, \bibinfo{author}{Yamakawa, Y.},
  \bibinfo{author}{Tanaka, Y.} \& \bibinfo{author}{Okamoto, Y.}
\newblock \bibinfo{title}{Line-node dirac semimetal and topological insulating
  phase in noncentrosymmetric pnictides {CaAgX} (x = p, as)}.
\newblock \emph{\bibinfo{journal}{Journal of the Physical Society of Japan}}
  \textbf{\bibinfo{volume}{85}}, \bibinfo{pages}{013708}
  (\bibinfo{year}{2016}).

\bibitem{IrF4_2016}
\bibinfo{author}{Bzdu{\v{s}}ek, T.}, \bibinfo{author}{Wu, Q.},
  \bibinfo{author}{Rüegg, A.}, \bibinfo{author}{Sigrist, M.} \&
  \bibinfo{author}{Soluyanov, A.~A.}
\newblock \bibinfo{title}{Nodal-chain metals}.
\newblock \emph{\bibinfo{journal}{Nature}} \textbf{\bibinfo{volume}{538}},
  \bibinfo{pages}{75--78} (\bibinfo{year}{2016}).

\bibitem{ZrSiS_2016}
\bibinfo{author}{Schoop, L.~M.} \emph{et~al.}
\newblock \bibinfo{title}{Dirac cone protected by non-symmorphic symmetry and
  three-dimensional dirac line node in {ZrSiS}}.
\newblock \emph{\bibinfo{journal}{Nature Communications}}
  \textbf{\bibinfo{volume}{7}}, \bibinfo{pages}{11696} (\bibinfo{year}{2016}).

\bibitem{Triple-2}
\bibinfo{author}{Zhu, Z.}, \bibinfo{author}{Winkler, G.~W.},
  \bibinfo{author}{Wu, Q.}, \bibinfo{author}{Li, J.} \&
  \bibinfo{author}{Soluyanov, A.~A.}
\newblock \bibinfo{title}{Triple point topological metals}.
\newblock \emph{\bibinfo{journal}{Physical Review X}}
  \textbf{\bibinfo{volume}{6}}, \bibinfo{pages}{031003} (\bibinfo{year}{2016}).

\bibitem{TopologicalQuantumChemisty}
\bibinfo{author}{Bradlyn, B.} \emph{et~al.}
\newblock \bibinfo{title}{Topological quantum chemistry}.
\newblock \emph{\bibinfo{journal}{Nature}} \textbf{\bibinfo{volume}{547}},
  \bibinfo{pages}{298--305} (\bibinfo{year}{2017}).
\newblock \urlprefix\url{https://doi.org/10.1038/nature23268}.

\bibitem{EightFold}
\bibinfo{author}{Bradlyn, B.} \emph{et~al.}
\newblock \bibinfo{title}{Beyond dirac and weyl fermions: Unconventional
  quasiparticles in conventional crystals}.
\newblock \emph{\bibinfo{journal}{Science}} \textbf{\bibinfo{volume}{353}},
  \bibinfo{pages}{aaf5037} (\bibinfo{year}{2016}).

\bibitem{TMDatabase}
\bibinfo{author}{Vergniory, M.~G.} \emph{et~al.}
\newblock \bibinfo{title}{A complete catalogue of high-quality topological
  materials}.
\newblock \emph{\bibinfo{journal}{Nature}} \textbf{\bibinfo{volume}{566}},
  \bibinfo{pages}{480--485} (\bibinfo{year}{2019}).

\bibitem{TMDatabasewebsite}
\bibinfo{title}{Topological material database}.
\newblock \urlprefix\url{https://topologicalquantumchemistry.org/#/}.

\bibitem{Liang_2014}
\bibinfo{author}{Liang, T.} \emph{et~al.}
\newblock \bibinfo{title}{Ultrahigh mobility and giant magnetoresistance in the
  dirac semimetal~cd3{A}s2}.
\newblock \emph{\bibinfo{journal}{Nature Materials}}
  \textbf{\bibinfo{volume}{14}}, \bibinfo{pages}{280--284}
  (\bibinfo{year}{2014}).
\newblock \urlprefix\url{http://dx.doi.org/10.1038/nmat4143}.

\bibitem{Narayanan_2015}
\bibinfo{author}{Narayanan, A.} \emph{et~al.}
\newblock \bibinfo{title}{Linear magnetoresistance caused by mobility
  fluctuations in n -doped cd 3~{A}s 2}.
\newblock \emph{\bibinfo{journal}{Phys. Rev. Lett.}}
  \textbf{\bibinfo{volume}{114}}, \bibinfo{pages}{117201}
  (\bibinfo{year}{2015}).
\newblock \urlprefix\url{http://dx.doi.org/10.1103/PhysRevLett.114.117201}.

\bibitem{Wang2013a}
\bibinfo{author}{Wang, Z.} \& \bibinfo{author}{Zhang, S.-C.}
\newblock \bibinfo{title}{Chiral anomaly, charge density waves, and axion
  strings from weyl semimetals}.
\newblock \emph{\bibinfo{journal}{Phys. Rev. B}} \textbf{\bibinfo{volume}{87}},
  \bibinfo{pages}{161107} (\bibinfo{year}{2013}).
\newblock \urlprefix\url{http://dx.doi.org/10.1103/PhysRevB.87.161107}.

\bibitem{Shekhar_2015}
\bibinfo{author}{Shekhar, C.} \emph{et~al.}
\newblock \bibinfo{title}{Extremely large magnetoresistance and ultrahigh
  mobility in the topological weyl semimetal candidate {NbP}}.
\newblock \emph{\bibinfo{journal}{Nat Phys}} \textbf{\bibinfo{volume}{11}},
  \bibinfo{pages}{645--649} (\bibinfo{year}{2015}).
\newblock \urlprefix\url{http://dx.doi.org/10.1038/nphys3372}.

\bibitem{ZrTe5_Chiral}
\bibinfo{author}{Li, Q.} \emph{et~al.}
\newblock \bibinfo{title}{Chiral magnetic effect in {ZrTe}5}.
\newblock \emph{\bibinfo{journal}{Nature Physics}}
  \textbf{\bibinfo{volume}{12}}, \bibinfo{pages}{550--554}
  (\bibinfo{year}{2016}).
\newblock \urlprefix\url{http://dx.doi.org/10.1038/nphys3648}.

\bibitem{TaAs-Jia}
\bibinfo{author}{Zhang, C.} \emph{et~al.}
\newblock \bibinfo{title}{Tantalum monoarsenide: an exotic compensated
  semimetal}.
\newblock \urlprefix\url{https://arxiv.org/abs/1502.00251}.

\bibitem{TaAs-Chen}
\bibinfo{author}{Huang, X.} \emph{et~al.}
\newblock \bibinfo{title}{Observation of the chiral-anomaly-induced negative
  magnetoresistance in 3d weyl semimetal {TaAs}}.
\newblock \emph{\bibinfo{journal}{Phys. Rev. X}} \textbf{\bibinfo{volume}{5}},
  \bibinfo{pages}{031023} (\bibinfo{year}{2015}).
\newblock \urlprefix\url{http://dx.doi.org/10.1103/PhysRevX.5.031023}.

\bibitem{NbAs-Luo}
\bibinfo{author}{Luo, Y.} \emph{et~al.}
\newblock \bibinfo{title}{Electron-hole compensation effect between
  topologically trivial electrons and nontrivial holes in {NbAs}}.
\newblock \emph{\bibinfo{journal}{Phys. Rev. B}} \textbf{\bibinfo{volume}{92}},
  \bibinfo{pages}{205134} (\bibinfo{year}{2015}).
\newblock \urlprefix\url{http://dx.doi.org/10.1103/PhysRevB.92.205134}.

\bibitem{ZrSiS-MR}
\bibinfo{author}{Ali, M.~N.} \emph{et~al.}
\newblock \bibinfo{title}{Butterfly magnetoresistance, quasi-2d dirac fermi
  surface and topological phase transition in {ZrSiS}}.
\newblock \emph{\bibinfo{journal}{Science Advances}}
  \textbf{\bibinfo{volume}{2}}, \bibinfo{pages}{e1601742}
  (\bibinfo{year}{2016}).

\bibitem{LaSb_Tafti}
\bibinfo{author}{Tafti, F.~F.}, \bibinfo{author}{Gibson, Q.~D.},
  \bibinfo{author}{Kushwaha, S.~K.}, \bibinfo{author}{Haldolaarachchige, N.} \&
  \bibinfo{author}{Cava, R.~J.}
\newblock \bibinfo{title}{Resistivity plateau and extreme magnetoresistance in
  {LaSb}}.
\newblock \emph{\bibinfo{journal}{Nature Physics}}
  \textbf{\bibinfo{volume}{12}}, \bibinfo{pages}{272--277}
  (\bibinfo{year}{2015}).
\newblock \urlprefix\url{http://dx.doi.org/10.1038/nphys3581}.

\bibitem{NbSb2-Wang}
\bibinfo{author}{Wang, K.}, \bibinfo{author}{Graf, D.}, \bibinfo{author}{Li,
  L.}, \bibinfo{author}{Wang, L.} \& \bibinfo{author}{Petrovic, C.}
\newblock \bibinfo{title}{Anisotropic giant magnetoresistance in {NbSb}2}.
\newblock \emph{\bibinfo{journal}{Sci. Rep.}} \textbf{\bibinfo{volume}{4}},
  \bibinfo{pages}{7328} (\bibinfo{year}{2014}).
\newblock \urlprefix\url{http://dx.doi.org/10.1038/srep07328}.

\bibitem{NbAs2_Shen}
\bibinfo{author}{Shen, B.}, \bibinfo{author}{Deng, X.},
  \bibinfo{author}{Kotliar, G.} \& \bibinfo{author}{Ni, N.}
\newblock \bibinfo{title}{Fermi surface topology and negative longitudinal
  magnetoresistance observed in the semimetal {NbAs} 2}.
\newblock \emph{\bibinfo{journal}{Physical Review B}}
  \textbf{\bibinfo{volume}{93}}, \bibinfo{pages}{195119}
  (\bibinfo{year}{2016}).
\newblock \urlprefix\url{http://dx.doi.org/10.1103/PhysRevB.93.195119}.

\bibitem{TaAs2_Luo}
\bibinfo{author}{Luo, Y.} \emph{et~al.}
\newblock \bibinfo{title}{Anomalous electronic structure and magnetoresistance
  in {TaAs}2}.
\newblock \emph{\bibinfo{journal}{Scientific Reports}}
  \textbf{\bibinfo{volume}{6}}, \bibinfo{pages}{27294} (\bibinfo{year}{2016}).
\newblock \urlprefix\url{http://dx.doi.org/10.1038/srep27294}.

\bibitem{LaBi_ARPES3}
\bibinfo{author}{Nayak, J.} \emph{et~al.}
\newblock \bibinfo{title}{Multiple dirac cones at the surface of the
  topological metal {LaBi}}.
\newblock \emph{\bibinfo{journal}{Nature Communications}}
  \textbf{\bibinfo{volume}{8}}, \bibinfo{pages}{13942} (\bibinfo{year}{2017}).

\bibitem{LaBi_ARPES2}
\bibinfo{author}{Wu, Y.} \emph{et~al.}
\newblock \bibinfo{title}{Asymmetric mass acquisition in {LaBi}: Topological
  semimetal candidate}.
\newblock \emph{\bibinfo{journal}{Physical Review B}}
  \textbf{\bibinfo{volume}{94}}, \bibinfo{pages}{081108}
  (\bibinfo{year}{2016}).

\bibitem{LaBi_ARPES1}
\bibinfo{author}{Lou, R.} \emph{et~al.}
\newblock \bibinfo{title}{Evidence of topological insulator state in the
  semimetal {LaBi}}.
\newblock \emph{\bibinfo{journal}{Physical Review B}}
  \textbf{\bibinfo{volume}{95}}, \bibinfo{pages}{115140}
  (\bibinfo{year}{2017}).

\bibitem{LaSb_ARPES1}
\bibinfo{author}{Zeng, L.-K.} \emph{et~al.}
\newblock \bibinfo{title}{Compensated semimetal {LaSb} with unsaturated
  magnetoresistance}.
\newblock \emph{\bibinfo{journal}{Physical Review Letters}}
  \textbf{\bibinfo{volume}{117}}, \bibinfo{pages}{127204}
  (\bibinfo{year}{2016}).

\bibitem{NbSb2_Z2}
\bibinfo{author}{Xu, C.} \emph{et~al.}
\newblock \bibinfo{title}{Electronic structures of transition metal dipnictides
  xpn2 (x=ta, nb; pn=p, as, sb)}.
\newblock \emph{\bibinfo{journal}{Physical Review B}}
  \textbf{\bibinfo{volume}{93}}, \bibinfo{pages}{195106}
  (\bibinfo{year}{2016}).

\bibitem{NbSb2_HiddenWeyl}
\bibinfo{author}{Gresch, D.}, \bibinfo{author}{Wu, Q.},
  \bibinfo{author}{Winkler, G.~W.} \& \bibinfo{author}{Soluyanov, A.~A.}
\newblock \bibinfo{title}{Hidden weyl points in centrosymmetric paramagnetic
  metals}.
\newblock \emph{\bibinfo{journal}{New Journal of Physics}}
  \textbf{\bibinfo{volume}{19}}, \bibinfo{pages}{035001}
  (\bibinfo{year}{2017}).

\bibitem{CoSb3-PRL}
\bibinfo{author}{Smith, J.~C.}, \bibinfo{author}{Banerjee, S.},
  \bibinfo{author}{Pardo, V.} \& \bibinfo{author}{Pickett, W.~E.}
\newblock \bibinfo{title}{Dirac point degenerate with massive bands at a
  topological quantum critical point}.
\newblock \emph{\bibinfo{journal}{Phys. Rev. Lett.}}
  \textbf{\bibinfo{volume}{106}}, \bibinfo{pages}{056401}
  (\bibinfo{year}{2011}).
\newblock \urlprefix\url{http://dx.doi.org/10.1103/PhysRevLett.106.056401}.

\bibitem{Tang_2015}
\bibinfo{author}{Tang, Y.} \emph{et~al.}
\newblock \bibinfo{title}{Convergence of multi-valley bands as the electronic
  origin of high thermoelectric performance in {CoSb}3 skutterudites}.
\newblock \emph{\bibinfo{journal}{Nature Materials}}
  \textbf{\bibinfo{volume}{14}}, \bibinfo{pages}{1223--1228}
  (\bibinfo{year}{2015}).
\newblock \urlprefix\url{http://dx.doi.org/10.1038/nmat4430}.

\bibitem{Singh_1994}
\bibinfo{author}{Singh, D.~J.} \& \bibinfo{author}{Pickett, W.~E.}
\newblock \bibinfo{title}{Skutterudite antimonides: Quasilinear bands and
  unusual transport}.
\newblock \emph{\bibinfo{journal}{Phys. Rev. B}} \textbf{\bibinfo{volume}{50}},
  \bibinfo{pages}{11235--11238} (\bibinfo{year}{1994}).
\newblock \urlprefix\url{http://dx.doi.org/10.1103/PhysRevB.50.11235}.

\bibitem{CoSb3-PRB}
\bibinfo{author}{Pardo, V.}, \bibinfo{author}{Smith, J.~C.} \&
  \bibinfo{author}{Pickett, W.~E.}
\newblock \bibinfo{title}{Linear bands, zero-momentum weyl semimetal, and
  topological transition in skutterudite-structure pnictides}.
\newblock \emph{\bibinfo{journal}{Phys. Rev. B}} \textbf{\bibinfo{volume}{85}},
  \bibinfo{pages}{214531} (\bibinfo{year}{2012}).
\newblock \urlprefix\url{http://dx.doi.org/10.1103/PhysRevB.85.214531}.

\bibitem{IrBi3}
\bibinfo{author}{Yang, M.} \& \bibinfo{author}{Liu, W.-M.}
\newblock \bibinfo{title}{The d-p band-inversion topological insulator in
  bismuth-based skutterudites}.
\newblock \emph{\bibinfo{journal}{Sci. Rep.}} \textbf{\bibinfo{volume}{4}}
  (\bibinfo{year}{2014}).
\newblock \urlprefix\url{http://dx.doi.org/10.1038/srep05131}.

\bibitem{Koga_2005}
\bibinfo{author}{Koga, K.}, \bibinfo{author}{Akai, K.},
  \bibinfo{author}{Oshiro, K.} \& \bibinfo{author}{Matsuura, M.}
\newblock \bibinfo{title}{Electronic structure and optical properties of binary
  skutterudite antimonides}.
\newblock \emph{\bibinfo{journal}{Physical Review B}}
  \textbf{\bibinfo{volume}{71}}, \bibinfo{pages}{155119}
  (\bibinfo{year}{2005}).

\bibitem{wzjcode}
\bibinfo{author}{Wang, Z.}
\newblock \bibinfo{title}{The open-source code vasp2trace and end-user button
  checktopologicalmat are available online at
  www.cryst.ehu.es/cryst/checktopologicalmat.}
\newblock \urlprefix\url{http://www.cryst.ehu.es/cryst/checktopologicalmat}.

\bibitem{OscillationBook}
\bibinfo{author}{Shoenberg, D.}
\newblock \emph{\bibinfo{title}{Magnetic oscillations in metals}}
  (\bibinfo{publisher}{Cambridge University Press ({CUP})},
  \bibinfo{year}{1984}).
\newblock \urlprefix\url{http://dx.doi.org/10.1017/CBO9780511897870}.

\bibitem{phase-shift1}
\bibinfo{author}{Murakawa, H.} \emph{et~al.}
\newblock \bibinfo{title}{Detection of berry's phase in a bulk rashba
  semiconductor}.
\newblock \emph{\bibinfo{journal}{Science}} \textbf{\bibinfo{volume}{342}},
  \bibinfo{pages}{1490--1493} (\bibinfo{year}{2013}).

\bibitem{phase-shift2}
\bibinfo{author}{He, L.} \emph{et~al.}
\newblock \bibinfo{title}{Quantum transport evidence for the three-dimensional
  dirac semimetal phase {inCd}3as2}.
\newblock \emph{\bibinfo{journal}{Physical Review Letters}}
  \textbf{\bibinfo{volume}{113}}, \bibinfo{pages}{246402}
  (\bibinfo{year}{2014}).

\bibitem{phase-shift3}
\bibinfo{author}{Luk'yanchuk, I.~A.} \& \bibinfo{author}{Kopelevich, Y.}
\newblock \bibinfo{title}{Dirac and normal fermions in graphite and graphene:
  Implications of the quantum hall effect}.
\newblock \emph{\bibinfo{journal}{Physical Review Letters}}
  \textbf{\bibinfo{volume}{97}} (\bibinfo{year}{2006}).

\bibitem{phase-shift4}
\bibinfo{author}{Luk'yanchuk, I.~A.} \& \bibinfo{author}{Kopelevich, Y.}
\newblock \bibinfo{title}{Phase analysis of quantum oscillations in graphite}.
\newblock \emph{\bibinfo{journal}{Physical Review Letters}}
  \textbf{\bibinfo{volume}{93}} (\bibinfo{year}{2004}).

\bibitem{InSb1}
\bibinfo{author}{Hu, J.} \& \bibinfo{author}{Rosenbaum, T.~F.}
\newblock \bibinfo{title}{Classical and quantum routes to linear
  magnetoresistance}.
\newblock \emph{\bibinfo{journal}{Nature Materials}}
  \textbf{\bibinfo{volume}{7}}, \bibinfo{pages}{697--700}
  (\bibinfo{year}{2008}).

\bibitem{InSb2}
\bibinfo{author}{Kim, Y.-S.}, \bibinfo{author}{Hummer, K.} \&
  \bibinfo{author}{Kresse, G.}
\newblock \bibinfo{title}{Accurate band structures and effective masses for
  {InP}, {InAs}, and {InSb} using hybrid functionals}.
\newblock \emph{\bibinfo{journal}{Physical Review B}}
  \textbf{\bibinfo{volume}{80}} (\bibinfo{year}{2009}).

\bibitem{Huang2015}
\bibinfo{author}{Huang, S.-M.} \emph{et~al.}
\newblock \bibinfo{title}{A weyl fermion semimetal with surface fermi arcs in
  the transition metal monopnictide {TaAs} class}.
\newblock \emph{\bibinfo{journal}{Nature Communications}}
  \textbf{\bibinfo{volume}{6}}, \bibinfo{pages}{7373} (\bibinfo{year}{2015}).
\newblock \urlprefix\url{http://dx.doi.org/10.1038/ncomms8373}.

\bibitem{Graphene_RMP}
\bibinfo{author}{Neto, A. H.~C.}, \bibinfo{author}{Guinea, F.},
  \bibinfo{author}{Peres, N. M.~R.}, \bibinfo{author}{Novoselov, K.~S.} \&
  \bibinfo{author}{Geim, A.~K.}
\newblock \bibinfo{title}{The electronic properties of graphene}.
\newblock \emph{\bibinfo{journal}{Reviews of Modern Physics}}
  \textbf{\bibinfo{volume}{81}}, \bibinfo{pages}{109--162}
  (\bibinfo{year}{2009}).
\newblock \urlprefix\url{http://dx.doi.org/10.1103/RevModPhys.81.109}.

\bibitem{Graphene_Novoselov_2005}
\bibinfo{author}{Novoselov, K.~S.} \emph{et~al.}
\newblock \bibinfo{title}{Two-dimensional gas of massless dirac fermions in
  graphene}.
\newblock \emph{\bibinfo{journal}{Nature}} \textbf{\bibinfo{volume}{438}},
  \bibinfo{pages}{197--200} (\bibinfo{year}{2005}).
\newblock \urlprefix\url{http://dx.doi.org/10.1038/nature04233}.

\bibitem{Bumps}
\bibinfo{author}{Kienzle, P.~A.}, \bibinfo{author}{Krycka, J.},
  \bibinfo{author}{Patel, N.} \& \bibinfo{author}{Sahin, I.}
\newblock \bibinfo{title}{Bumps: Curve fitting and uncertainty analysis
  (version 0.7.6) [computer software]. college park, md: University of
  maryland} (\bibinfo{year}{2018}).
\newblock \urlprefix\url{https://doi.org/10.5281/zenodo.1249714}.

\bibitem{ando}
\bibinfo{author}{Taskin, A.~A.} \& \bibinfo{author}{Ando, Y.}
\newblock \bibinfo{title}{Berry phase of nonideal dirac fermions in topological
  insulators}.
\newblock \emph{\bibinfo{journal}{Physical Review B}}
  \textbf{\bibinfo{volume}{84}} (\bibinfo{year}{2011}).

\bibitem{wien2k1}
\bibinfo{author}{Weinert, M.}, \bibinfo{author}{Wimmer, E.} \&
  \bibinfo{author}{Freeman, A.~J.}
\newblock \bibinfo{title}{Total-energy all-electron density functional method
  for bulk solids and surfaces}.
\newblock \emph{\bibinfo{journal}{Phys. Rev. B}} \textbf{\bibinfo{volume}{26}},
  \bibinfo{pages}{4571--4578} (\bibinfo{year}{1982}).
\newblock \urlprefix\url{http://dx.doi.org/10.1103/PhysRevB.26.4571}.

\bibitem{wien2k2}
\bibinfo{author}{Blaha, P.}, \bibinfo{author}{Schwarz, K.},
  \bibinfo{author}{Madsen, G. K.~H.}, \bibinfo{author}{Kvasnicka, D.} \&
  \bibinfo{author}{Luitz, J.}
\newblock \emph{\bibinfo{title}{An Augmented PlaneWave + Local Orbitals Program
  for Calculating Crystal Properties}} (\bibinfo{publisher}{Karlheinz Schwarz,
  Techn. Universitat Wien, Austria}, \bibinfo{year}{2001}).
\newblock \urlprefix\url{http://www.wien2k.at/}.

\bibitem{gga}
\bibinfo{author}{Perdew, J.~P.}, \bibinfo{author}{Burke, K.} \&
  \bibinfo{author}{Ernzerhof, M.}
\newblock \bibinfo{title}{Generalized gradient approximation made simple}.
\newblock \emph{\bibinfo{journal}{Phys. Rev. Lett.}}
  \textbf{\bibinfo{volume}{77}}, \bibinfo{pages}{3865--3868}
  (\bibinfo{year}{1996}).
\newblock \urlprefix\url{http://dx.doi.org/10.1103/PhysRevLett.77.3865}.

\bibitem{mBJ}
\bibinfo{author}{Tran, F.} \& \bibinfo{author}{Blaha, P.}
\newblock \bibinfo{title}{Accurate band gaps of semiconductors and insulators
  with a semilocal exchange-correlation potential}.
\newblock \emph{\bibinfo{journal}{Physical Review Letters}}
  \textbf{\bibinfo{volume}{102}}, \bibinfo{pages}{226401}
  (\bibinfo{year}{2009}).

\bibitem{server}
\bibinfo{author}{Cano, J.} \emph{et~al.}
\newblock \bibinfo{title}{Building blocks of topological quantum chemistry:
  Elementary band representations}.
\newblock \emph{\bibinfo{journal}{Physical Review B}}
  \textbf{\bibinfo{volume}{97}}, \bibinfo{pages}{035139}
  (\bibinfo{year}{2018}).
\newblock \eprint{http://arxiv.org/abs/1709.01935v3}.

\bibitem{crystallographicserver}
\bibinfo{title}{Bilbao crystallographic server}.
\newblock \urlprefix\url{http://www.cryst.ehu.es/cryst/bandrep}.

\end{thebibliography}


\end{document}